\documentclass{article}

\usepackage{arxiv}

\usepackage[utf8]{inputenc} 
\usepackage[T1]{fontenc}    
\usepackage{hyperref}       
\usepackage{url}            
\usepackage{booktabs}       
\usepackage{amsfonts}       
\usepackage{nicefrac}       
\usepackage{microtype}      
\usepackage{lipsum}		
\usepackage{graphicx}
\usepackage[square,sort,comma,numbers]{natbib}
\usepackage{doi}
\usepackage{microtype}      
\usepackage{lipsum}
\usepackage{times}
\usepackage{subcaption}
\usepackage{amsmath, amssymb}
\usepackage{xcolor}
\usepackage{graphicx}
\usepackage{caption, subcaption}
\usepackage{bbm}
\usepackage{siunitx} 
\usepackage{pgfplotstable} 
\usepackage{multirow}
\usepackage{color, colortbl}
\usepackage[utf8]{inputenc}
\usepackage[english]{babel}
\usepackage{placeins}
\usepackage{float}

\title{Nonparametric scanning for nonrandom missing data with continuous instrumental variables}

\date{November 05, 2021}	

\author{ Arkaprabha Ganguli\\
	Department of Statistics and Probability\\
	Michigan State University\\
	East Lansing,  Michigan 48824, U.S.A. \\
	\texttt{gangulia@msu.edu} \\
	\And
	David Todem \thanks{Corresponding author : David Todem}\\ 
	Department of Epidemiology and Biostatistics\\
	Michigan State University\\
	East Lansing,  Michigan 48824, U.S.A. \\
	\texttt{todem@msu.edu} \\
}

\renewcommand{\shorttitle}{Nonparametric Scanning For Nonrandom Missing Data }

\hypersetup{
pdftitle={Nonparametric scanning for nonrandom missing data with continuous instrumental variables},
pdfauthor={Arkaprabha Ganguli, David Todem},
pdfkeywords={Continuous instrument variable;  Dichotomizations; Empirical processes; Exponential tilting; Identifiability; Kernel regression.},
}
\newtheorem{theorem}{Theorem}
\newtheorem{condition}{Condition}

\begin{document}
\maketitle

\begin{abstract}
This article introduces a new instrumental variable approach for estimating unknown population parameters with data having nonrandom missing values. With coarse and discrete instruments, Shao and Wang (2016) proposed a semiparametric method that  uses the added information to identify the tilting parameter from the missing data propensity model. A naive application of this idea to continuous instruments through arbitrary discretizations is apt  to be inefficient, and maybe questionable in some settings. We propose a nonparametric method not requiring arbitrary discretizations  but involves  scanning
over continuous dichotomizations of the instrument; and  combining scan statistics to estimate the unknown parameters via  weighted integration.
 We establish the asymptotic normality of the proposed integrated estimator and that of the underlying scan processes uniformly across the instrument sample space. Simulation studies and the analysis of a real data set demonstrate the gains of the methodology over procedures that rely either on arbitrary discretizations or moments of the instrument.
\end{abstract}


\section{Introduction}\label{Introduction}

Analysis of incomplete data poses a recurring challenge in statistics. It is well-established that inferences may be misleading if the working model assumptions are not consistent with the underlying missing data mechanism (Molenberghs \emph{et al.}, 1997\nocite{MOLENBERGHS:KENWARD:LESAFFRE:1997}).  Existing  strategies oftentimes used to account for some degree of selection bias due to nonresponse, typically rely on the  assumption that missingness can effectively be rendered independent of the outcome, upon conditioning on a sufficiently rich set of observed variables  (Robins \emph{et al.}, 1995\nocite{robins:rotnitzky:zhao:1994};
 Tsiatis \& Davidian, 2007\nocite{tsiatis2007}; Rubin, 1987\nocite{rubin:1987}; Little \& Rubin, 2002; and Kim \& Shao, 2021)\nocite{Litt:Rubi:stat:2002}\nocite{kim:shao:2021}. This assumption is most often made in practice, and formally entails a missing at random (MAR) data mechanism. Without imposing additional conditions, such a mechanism  is strictly untestable in view of observed data. Consequently, it is often prudent to conservatively assume that systematic differences between respondents and nonrespondents persist even after covariate adjustment, consistent with nonrandom nonresponse mechanism (Rotnitzky \emph{et al.}, 2001; Linero, 2017\nocite{linero:2017}; and Yang \emph{et al.}, 2019)\nocite{yang:2019}\nocite{Scha:Rotn:Robi:adju:1999}\nocite{Rotn:Scha:Su:Robi:meth:2001}.

Identifiability issues commonly arise with nonrandom missing data, where the parameters of
the model for the missingness may not be jointly identifiable with those from the measurement model using only the observed data (Scharfstein \emph{et al.}, 1999; Todem \emph{et al.}, 2010; Zhao and Shao, 2015\nocite{zhao:shao:2015})\nocite{Scha:Rotn:Robi:adju:1999}\nocite{Todem:2009}.
Analytic strategies addressing these concerns include
methods that rely on identification on parametric restrictions which may be subject to misspecification (Wu
\& Carroll, 1988\nocite{wu:carroll:1988}; Molenberghs \emph{et al.}, 1997\nocite{MOLENBERGHS:KENWARD:LESAFFRE:1997};
Troxel  \emph{et al.}, 1998\nocite{Trox:Lips:Harr:marg:1998} and Wang \emph{et al.}, 2014\nocite{wang:2014}).

A more recent method that has been shown to be essential for the model identification is the use of an instrumental variable (IV). At least two versions of this approach have appeared in the literature. While the first version treats the IV as being independent of the missing data indicator given the response (Shao \& Wang, 2016\nocite{shao:wang:2016}), the second version rather treats the IV as being independent of the outcome given the missing data indicator (Tchetgen Tchetgen and Wirth, 2017\nocite{tchetgen:wirth:2017}). Although these formulations may have different substantive  interpretations, they are fundamentally equivalent from a distributional viewpoint. In fact, one can argue that they are closely connected to the selection and pattern mixture model taxonomy. For this reason, we will adopt, in  the spirit of Shao \& Wang (2016), the first IV representation which allows a direct characterization of parameters indexing the law of data given covariates.

To fix ideas, let $y$ denote a univariate outcome subject to nonresponse, and $\delta$ the associated response status indicator with $\delta=1$ indicating an observed value. Let $z$ be  a nonresponse IV, and $u$ an auxiliary variable, both assumed to be continuous and fully observed. As such, $z$ is assumed to be associated with $\delta$ only through its dependence on $y$, such that $z$ and $\delta$ are conditionally independent given $y$.  Consequently, $z$ must predict a person's propensity to have an observed outcome through its direct influence on the outcome itself; that is $\mbox{pr}(\delta=1\mid y, u, z)=\mbox{pr}(\delta=1\mid y, u)$. Letting $x=(u, z)$ and $\pi(y,x)=\mbox{pr}(\delta=1\mid y, x)$ and
following the formulation of Kim \& Yu (2011) \nocite{kim:yu:2011}, we consider an exponential tilting model of the form,
\begin{equation} 
\pi(y,x)=\left[ 1+ \exp\{g(u)+\gamma y\} \right]^{-1},
\end{equation}\label{tilting}
where $\gamma$ is the titling parameter measuring the extent of deviation of $\delta$ from  MAR, $g$ is an unknown and completely unspecified function of $u$. Unlike other approaches that rely on either fixing $\gamma$ to some known value or estimating $\gamma$ using external data (e.g., Kim and Yu, 2011; Zhao et al, 2013 and Tang et al, 2014), added information from IVs can be used to formulate additional estimating equations for $\gamma$ while profiling $g$. With discrete IVs, Shao \& Wang (2016) showed that the profiled estimating equations can be used to consistently estimate $\gamma$. A straightforward application of this idea to continuous IVs via arbitrary discretizations is apt to be suboptimal. Because of its dependency on  thresholds,  this approach may fail to provide a unique answer to the substantive question in some  applications. It may further be subject to the theoretical complications arising with the data-dependent threshold placement. A comprehensive discussion on the danger of discretizing continuous variables can be found elsewhere (e.g., Altman, 1998\nocite{Altman:1998}; Royston \emph{et al.}, 2006; Farewell {\emph{et al.}}, 2004 and Peng \& Fine, 2008\nocite{peng:fine:2008})\nocite{Royston:Altman:Sauerbrei:2006}\nocite{farewell:2004}.

 In this paper, we propose a novel and flexible approach that does not rely on arbitrary discretizations but involves scanning through continuous thresholding of the instrument and combining scan statistics to estimate the unknown parameter through weighted integration. The key to effective inference under continuous thresholding lies in its ability to improve efficiency, especially when there are local dependencies between $z$ and $\delta$ involving a small portion of the sample. An interesting feature of the scanning approach is that the efficient gain  extends naturally to settings involving multiple instruments, albeit a heavy  computational cost. Numerical simulations and an analysis of data generated from an income panel study demonstrate the gains of the methodology over procedures that rely either on arbitrary discretizations or moments of the instrument. Technical details are given in the Supplementary Material.


\section{The scanning method}

Suppose one is interested in estimating $\theta$, a $q$-dimensional parameter vector from the law $f(y\mid x)$ that uniquely solves $E\{\eta(y,x, \theta)\mid x \}=0$, for a generic data function $\eta(.)$. To this end, assume that an independent and identically distributed sample $\{y_i, x_i, \delta_i\}_{i=1}^{n}$ of $\{y,x,\delta\}$ is readily available. As alluded to in the introduction, we exploit the information contained in instrument $z$ to identify and estimate the tilting parameter $\gamma$ from the propensity  model in (\ref{tilting}). Let $\Xi\subseteq \mathbb{R}$ be the support of $z$, and $\Xi^o$ its interior. A naive strategy is to discretize $z$ using say $p\ge 1$ finite arbitrary thresholds $\xi_1 < \xi_2<\cdots < \xi_p$  of $\Xi^o$ to construct $p+1$ instrumental estimating functions,
\begin{equation}\label{estimating}
M_{\xi_{j}}(y,u,\delta; g,\gamma)= I\{\xi_{j-1} < z \leq \xi_{j}\}\left\{ {\delta \over \pi(y,u;g,\gamma)} -1 \right\},\; j=1, \cdots, p+1.
\end{equation}
Here we set $\xi_0=-\infty$ and  $\xi_{p+1}=+\infty$, and $I\{A\}$ is the indicator function for event $A$. Letting  $M_\xi$ be a column vector made of functions $\{M_{\xi_{j}}\}_{j=1}^{p+1}$
  obtained through stratification, we define the instrumental estimating equation,
\begin{equation} \label{est-eqn}
E\{M_\xi(y,u,\delta; g,\gamma)\mid u \}=0.
\end{equation}
This estimating function is made possible because of conditional independence between $z$ and $\delta$ given  $y$. For a fixed $\gamma$, let $g_{\gamma}$ be the function defined such that,
\[
    \exp\{g_\gamma(u)\}= \frac{E\{(1-\delta)\mid u\}}{E\{\delta \exp(\gamma y)\mid u\}}.
\]
Let $\gamma_0$ be the true value of $\gamma$ that solves (\ref{est-eqn}). From Equation $E\{\mathbbm{1}^{\prime} M_\xi(y,u,\delta; g,\gamma)\mid u \}=0,$ where $\mathbbm{1}$ is a $(p+1)$-column vector of 1, it can be readily shown that,
\begin{equation}\label{g-profile}
\exp\{g(u)\}={E\{(1-\delta)\mid u \} \over E\{\delta\exp(\gamma_0 y)\mid u \} },
\end{equation}
such that, $g={g}_{\gamma_0}$. Because of the conditional expectations in (\ref{g-profile}), a kernel method may be used to estimate $g(u)={g}_{\gamma_0}(u)$ provided $\gamma_0$ the true $\gamma$ is known or estimated from historical data. Because prior information on $\gamma_0$ is oftentimes unavailable in real applications, one may rely on $z$ for its estimation. First, we profile $g_\gamma(u)$ for fixed $\gamma$ as,
\[ \exp\{\hat{g}_\gamma(u)\}={{\sum_{i=1}^n}(1-\delta_i)K_h(u-u_i)\over {{\sum_{i=1}^n}\delta_i \exp\{\gamma y_i\} K_h(u-u_i)}}, \]
 where $K_h(.)=K(.)/h$, with $K(.)$ being a symmetric kernel function and $h$ a bandwidth. Estimation of $\gamma$ then follows by solving the  sample version of equation (\ref{est-eqn}),
\begin{equation}\label{est_gamma}
{1 \over n}\sum_{i=1}^n M_\xi(y_i,u_i,\delta_i; \hat{g}_\gamma,\gamma)=0.
\end{equation}
 Upon estimating $\gamma$ by $\hat{\gamma}$ and $g_{\gamma_0}$ with $\hat{g}_{\hat{\gamma}}$, one may apply the inverse propensity weighting approach to estimate $\theta$. Specifically, an estimator of $\theta$ may be obtained by solving the equation,
\begin{equation}\label{theta_initial} \sum_{i=1}^n {\delta_i\eta(y_i,x_i, \theta)\over \pi(y_i,u_i;\hat{g}_{\hat{\gamma}},\hat{\gamma})}=0. \end{equation}
As  $n \rightarrow \infty$, it can be shown under basic regularity conditions that the resulting estimator of $\theta$ converges asymptotically  at the $\sqrt{n}$ rate to a normal distribution, despite $g(u)$ being nonparametric. An even stronger asymptotic result can be established for $\hat{\gamma}$ (Shao \& Wang, 2016).

Because this estimation approach relies on arbitrary choices of thresholds and partial information from the instrument $z$, $I\{\xi_{j-1} < z \leq \xi_{j}\},\; j=1, \cdots, p+1$, the resulting estimators of $\gamma$ and $\theta$ are likely to be suboptimal. This deficiency may be alleviated with a proper choice of $\{\xi_j\}_{j=1}^p$, which is generally a daunting task in practice without knowing the true data generating process. A natural and conservative strategy avoiding arbitrary choices of thresholds, is to vary the thresholds continuously over the sample space and accumulate results. Our idea then proceeds through scanning via continuous dichotomizations, which is computationally very efficient while maintaining information contained in the instrument $z$.  This desirable feature of continuous dichotomizations relies on the property that the Borel $\sigma$-$algebra$ for the real line $\mathbb{R}$ is generating by half-infinite intervals and by bounded intervals; that is  $\sigma(\{(-\infty, b]: b \in \mathbb{R}\})=\sigma(\{(a, b]: a, b \in \mathbb{R}\})$;  thus $\sigma(\{(-\infty, b]\cap \Xi: b \in \mathbb{R}\})=\sigma(\{(a, b]\cap \Xi: a, b \in \mathbb{R}\})$.

Operationally, we set  $p=1$ and  regard estimators of $\gamma$ and $\theta$ in (\ref{est_gamma}) and (\ref{theta_initial}) as processes in single threshold $\xi$, which we denote $\tilde{\gamma}_\xi$ and $\tilde{\theta}_\xi$ to highlight this dependence. From these processes in $\xi$, we estimate $\theta$ as,
\[
\hat{\theta}=\mbox{argmin}_{\theta \in \Theta} \int_{\Xi^*} (\tilde{\theta}_\xi-\theta)^{T} {W}_\xi (\tilde{\theta}_\xi-\theta)d\xi,
\]
where ${\Xi^*} \subset {\Xi^0}$ with general requirements specified in \S \ref{theory}, ${W}_\xi$ is a positive definite matrix of weights  converging to $w_\xi$ uniformly in $\xi$. An estimator of $\theta$ is,
\begin{equation} \label{final-estimator}
\hat{\theta}=\left\{{\int_{\Xi^*} {W}_\xi d\xi}\right\}^{-1}{\int_{\Xi^*} \tilde{\theta}_\xi {W}_\xi d\xi}.
 \end{equation}
 Weights ${W}_\xi$ are chosen  to ensure the stability of $\hat{\theta}$, with special cases being  $\tilde{V}(\tilde{\theta}_\xi)^{-1}$ the inverse of the estimated variance-covariance of $\tilde{\theta}_\xi$, and $\{1/\tilde{p}_{\xi}+ 1/(1-\tilde{p}_{\xi})\}^{-1}I_{q}=\tilde{p}_{\xi}(1-\tilde{p}_{\xi})I_{q}$; $\tilde{p}_{\xi}$ is the empirical version of $Pr(z\leq \xi)$, and $I_{q}$ the identity matrix of order $q$.
While $\tilde{V}(\tilde{\theta}_\xi)^{-1}$ puts more weights on estimates that are less  variable, $p_{\xi}(1-p_{\xi})I_{q}$ weighs more scans toward  the median of $z$.

It is worth noting that while the current presentation focuses primarily on continuous instruments, the proposed method is generally applicable to IVs with an intrinsic order. A good example of such instruments is the class of ordinal categorical variables where the underlying ranking is  automatically incorporated into the estimation. For such instruments, one would argue that the proposed scanning approach will generally outperform the method proposed by Shao \& Wang (2016) which neglects the natural order of the instrument.

A competitive approach to the scanning method  with continuous instruments, is the moment-based procedure, consisting of estimating $\gamma$ via the equation,
\begin{equation}\label{estimating}
E\left\{ \left. z^{\ell}\left\{ {\delta \over \pi(y,u;g,\gamma)} -1 \right\}   \right| u \right\}=0, \; \ell=1, \cdots, L.
\end{equation}
As we shall see hereafter, this approach has limited applications when $z$ is sparse albeit being quantitative. Furthermore,  it is unclear how to choose $L$ or apply this approach to ordinal categorical instruments for which  the scanning approach remains applicable.

\section{Theory}\label{theory}

Although the idea of scanning and integrating may appear intuitive, formally establishing the large sample properties of the resulting estimator is not trivial owing to the dependency among scan statistics. Consequently, a careful study using empirical process theory may be invoked for asymptotics. To this end, we adopt the primitive requirements of Shao \& Wang (2016) pertaining to  $\tilde{\gamma}_\xi$ and $\tilde{\theta}_\xi$ for each fixed $\xi  \in {\Xi}^{*}$,  which we regard as pointwise conditions. These weaker requirements are then strengthened with higher level conditions to guarantee the uniform convergence of empirical processes in $\xi$, $\tilde{\gamma}_\xi$ and $\tilde{\theta}_\xi$.

The following theorems establish key asymptotic results: uniform consistency and asymptotic normality of $\hat{\gamma}_\xi$; uniform  asymptotic distribution of the stochastic process $\hat{\theta}_\xi$ and finally the asymptotic normality of our proposed estimator $\hat{\theta}$.
 It is worth noting that $\hat{\gamma}_\xi$ for each fixed  $\xi$, falls into the realm of ‘Semiparametric-M’ estimators for which general pointwise asymptotic results have been established (e.g., Khan \& Powell, 2001)\nocite{khan:2001}. Theorems \ref{thm1} and \ref{thm2} give a stronger uniform results across the threshold space.

\begin{theorem}\label{thm1}
Assume $Pr(z\leq \inf{\Xi^*})>0$, $Pr(z\geq \sup{\Xi^*})>0$ and Conditions 1-5 in Appendix hold. Let $\gamma_{0\xi}$ be the unique solution to $m_\xi (g_\gamma, \gamma)=0$, with $m_\xi (g_\gamma, \gamma)=E\{M_\xi(y,u,\delta; g,\gamma)\mid u \}$. If  $\sup\limits_{\gamma \in \Lambda,\; \xi \in \Xi^*} ||m_\xi (g_\gamma, \gamma)||<\infty$,  then $\sup\limits_{\xi \in \Xi^*} |\hat{\gamma}_\xi - \gamma_{0\xi}| \overset{P}{\to} 0$, as $n \to \infty$.
\end{theorem}

\begin{theorem}  \label{thm2}
Assume $Pr(z\leq \inf{\Xi^*})>0$, $Pr(z\geq \sup{\Xi^*})>0$ and Conditions 1-7 in Appendix hold. Then as $n \to \infty$, the random process $n^{1/2}(\hat{\gamma}_\xi - \gamma_{0\xi})$  in $\xi$ converges in distribution to a zero mean  Gaussian process with some covariance kernel $\sigma_0(\cdot,\cdot)$.
\end{theorem}

 Unlike $\hat{\gamma}_\xi$ which has a converging point dependent on $\xi$ both in probability and distribution, we show that $\hat{\theta}_\xi$ only converges in distribution and the centering point does not depend on $\xi$. For this reason, one would expect the scanning process to have an impact on precision but not on bias, at least in large samples.

\begin{theorem} \label{thm3}
Assume $Pr(z\leq \inf{\Xi^*})>0$, $Pr(z\geq \sup{\Xi^*})>0$ and Conditions 1-7 in Appendix hold. Then as $n \to \infty$, the random process $n^{1/2}(\hat{\theta}_\xi - \theta)$ in $\xi$, converges in distribution to a zero mean tight Gaussian process $\cal G$ with some covariance kernel $\sigma_1(\cdot,\cdot)$.
\end{theorem}

To prove Theorem \ref{thm3}, pertaining to the convergence of $v_{n}(\xi)=n^{1/2}(\hat{\theta}_\xi - \theta)$,  one needs to show (Dudley, 1978)\nocite{dudley:1978}; i) the sequence $(v_{n}(\xi_1), v_{n}(\xi_2),\cdots, v_{n}(\xi_k))$  converges in distribution to a centered multivariate normal distribution for every finite set of points $\{\xi_j\}_{j=1}^k$ of $\Xi$; and  ii) the asymptotic equicontinuity condition
 \[\forall \epsilon >0,   \lim\limits_{\delta \downarrow 0} \limsup\limits_{n\to\infty} P[\sup\limits_{|\xi-\xi'|<\delta} \parallel v_{n}(\xi)-v_{n}(\xi')\parallel>\epsilon]=0.\]

\begin{theorem}\label{thm4}
Under Conditions 1-8 specified in Appendix and finite variance assumption specified in the Supplementary Material, $n^{1/2} (\hat{\theta}-\theta)$ the scaled version of the integrated estimator $\hat{\theta}$ converges in distribution to a centered normal distribution $N(0, \Sigma)$, as $n \to \infty$.
\end{theorem}
The asymptotic variance-covariance $\Sigma$  has a very complicated form and may require resampling techniques for estimation. Basic regularity conditions, including those pertaining to the kernel estimation of the unspecified function $g$, are relegated to the Appendix. Proofs of all theorems except theorem \ref{thm2}, which follows directly from that of Shao \& Wang (2016), are provided in the Supplementary Material.

\section{Simulation Study}
\label{sec:Simulation Study}

We compare the small sample performance of the scanning procedure with that of competing methods for the estimation of the overall mean $\mu=E\{y\}$.
Specifically, in addition to the scanning-based estimator $\hat{\mu}$ of $\mu$, our investigation focuses on the following alternatives:
\begin{itemize}
  \item $\tilde{\mu}_{\xi_{\alpha}}$, the estimator with a single fixed threshold at quantile $\xi_{\alpha}$
  such that $P(z\leq \xi_{\alpha})=\alpha$, $\alpha \in \{0.2, 0.4, 0.6 \}$, and with a slight abuse of notation we simply write $\tilde{\mu}_{\alpha}$,
  \item $\tilde{\mu}_{\{0.2,0.4\}}$ and $\tilde{\mu}_{\{0.4,0.6\}}$, the estimator with two fixed thresholds at quantiles $\{0.2,0.4\}$ and $\{0.4,0.6\}$  of $ z$,
    \item $\Bar{y}=\sum_{i=1}^{n} y_i/n$, the sample mean of $y$ when there are no missing data,
  \item $\Bar{y}_{obs}=\sum_{i=1}^{n} \delta_iy_i/\sum_{i=1}^{n} \delta_i$, the sample mean of the observed $y_i$, complete case analysis,
  \item $\breve{\mu}_{\gamma}$ the estimator of Kim \& Yu (2011), with known tilting parameter $\gamma$ taking values $\gamma_0$ the true parameter, $\gamma_0/n$ the wrong tilting parameter, or  $0$ for the MAR assumption,
  \item $\check{\mu}$ the moment-based estimator with $L=1$.
\end{itemize}

The estimator $\hat{\mu}$ of $\mu$ is obtained by setting $\Xi^*=(\xi_{0.1}, \xi_{0.9})$, and weighing the scan statistics $\tilde{\mu}_{\xi}= \sum_{i=1}^n {\delta_i y_i/\pi(y_i,u_i;\hat{g}_{\hat{\gamma}_\xi},\hat{\gamma}_\xi)}$ by the inverse of their estimated variance. That is,
\[\hat{\mu}=\left\{{\int_{\Xi^*} {\tilde{V}(\tilde{\mu}_\xi)}^{-1}d\xi}\right\}^{-1} {\int_{\Xi^*} \tilde{\mu}_\xi {\tilde{V}(\tilde{\mu}_\xi)}^{-1}d\xi}.\]

 Estimators $\tilde{\mu}_{0.2}, \tilde{\mu}_{0.4}, \tilde{\mu}_{0.6}$, $\tilde{\mu}_{\{0.2,0.4\}}$ and $\tilde{\mu}_{\{0.4,0.6\}}$ are based on the approach in Shao \& Wang (2016) for categorical instruments with two  and three categories. The unspecified function $g(.)$ in the exponential tilting model is estimated using a nonparametric Gaussian Kernel $K(u)= exp(-u^2/2)/(2\pi)^{1/2}$ (Gasser \& M$\ddot{\rm u}$ller, 1979)\nocite{gasser:muller:1979}. The bandwidths are selected adaptively to the sample stratification for  estimation involving both fixed and continuous discretizations. For example with  single thresholds, the bandwidths are selected depending on $\xi$, i.e., $h_{n,\xi} = 1.5 \hat{\sigma}_{u_\xi}n_\xi^{-1/3}$ where $\hat{\sigma}_{u_\xi}$ is the standard deviation  of the observed variable $u_i$ with $z_i \leq \xi$; and $n_{\xi}=\sum_{i=1}^n I(z_i \leq \xi)$ is the stratum sample size.

Because the proposed method is applicable to instruments with an intrinsic order, the investigation is conducted for  continuous and ordinal categorical instruments. For  each data setting, we set the sample size to 200 and 500, and use 1000 Monte Carlo replicates to estimate the method's relative bias, mean square errors, standard errors and coverage probability of confidence intervals (generated using normal approximations with bootstrapped standard errors).

\subsection{Using continuous instrumental variable}\label{continuous_z}

We consider the following hierarchical structure for the full complete data:
\[z\sim N(0,1),\; u \mid \{z\} \sim N(z,1),\;  y \mid \{u, z\} \sim N(E\{y | u,z\}, 1).\]
Here $E\{y | u,z\}=\left[\sin(u^3) + 10u^2 \right] I\{z\leq 0.84\} + \left[40 + \sin(u^5)+10u \right] I\{z > 0.84\}$, which by averaging across $\{u,z\}$ gives  the unconditional mean $\mu=24.4$. Upon generating $\{z,u,y\}$, we generate the missingness indicator $\delta$ from Bernoulli distribution with probability $\pi$. Four MNAR  models ($M_1-M_4$), and one MAR model ($M_5$) are entertained, resulting in the overall missing data rate of about 30\%:\\
$M_1:\;$ $\pi=1/\{1+exp(\alpha+\beta u+\gamma y)\}$ with ($\alpha, \beta, \gamma)=(2.8,-0.3,-0.6)$;\\
$M_2:\;$ $\pi=1/\{1+exp(\alpha+\beta sin(u)+\gamma y)\}$ with ($\alpha, \beta, \gamma)=(2.75,-0.2,-0.6)$;\\
$M_3:\;$ $\pi=1/\{1+exp(\alpha+\beta u^2+\gamma y)\}$ with ($\alpha, \beta, \gamma)=(2.8,-0.3,-0.6)$;\\
$M_4:\;$ $\pi=1/\{1+exp(\alpha+\beta exp(u)+\gamma y)\}$ with ($\alpha, \beta, \gamma)=(3,-0.3,-0.6)$;\\
$M_5:\;$ $\pi=1/\{1+exp(\alpha+\beta exp(u))\}$ with ($\alpha, \beta)=(-0.9,-0.3)$;

Results from Tables \ref{tab:tbl1} and \ref{tab:tbl2} show that $\Bar{y}_{obs}$ is biased and does not have good coverage probability  under all missing data models considered. No other estimation method, except the benchmark $\bar{y}$, dominates the others uniformly across all data configurations, both in terms of bias and efficiency. The estimator $\breve{\mu}_{\gamma}$ which also serves as a benchmark when the true $\gamma$ is used, also has superior performance, both under the four MNAR and the MAR models. As expected, the naive estimators based on fixed single thresholds appear to have varying performance in realistic samples, that stabilizes with further discretizations. When pitched against the naive estimators, the scanning approach behaves robustly across most scenarios. This is consistent with early theoretical results that the scanning method enjoys some form of optimality as it accumulates information across all cutpoints. It is worth noting that when the 60$th$ percentile is used, $\tilde{\mu}_{0.6}$ outperforms $\hat{\mu}$ under all MNAR models. Because finding such an optimal cutpoint is a daunting task in applications, continuous discretizing appears to be the most conservative strategy. This is especially true if there are local dependencies between $z$ and $\delta$ involving a small portion of the sample.  As expected, $\breve{\mu}_{\gamma}$ performs poorly when a wrong titling parameter is assumed. For example, $\breve{\mu}_{0}$ and $\breve{\mu}_{\gamma_0/n}$ perform poorly under the MNAR models $M_1-M_4$.

It is also interesting to note that $\hat{\mu}$ outperforms $\check{\mu}$ under $M_5$, but the underlying reason is unclear. Further simulations are conducted to compare $\hat{\mu}$ and $\check{\mu}$ in settings involving an instrument with a two-component mixture distribution (results relegated to Supplementary Material). Essentially under zero-inflated continuous data, the moment-based estimator performs poorly regardless of the degree of the moment $L$ being chosen. This simulation result provides an excellent example of the limitation of this method in settings involving sparse data on the instrument.

\begin{table}[t!]
\def~{\hphantom{0}}
\centering
\caption{Relative biases (RBs), Mean square errors (MSEs) for n=200 and 500; all RBs are multiplied by 100}{%
\begin{tabular}{lcrrrrrrrrrr}
\\
& &  \multicolumn{10}{c}{Model}\\
&  & \multicolumn{2}{c}{$M_1$} & \multicolumn{2}{c}{$M_2$} & \multicolumn{2}{c}{$M_3$} & \multicolumn{2}{c}{$M_4$} & \multicolumn{2}{c}{$M_5$}   \\
Estimator & Quantity  \hspace{0.1in} $n=$ \hspace{-0.1in} & 200 & 500 & 200 & 500 & 200 & 500 & 200 & 500 & 200 & 500  \\
\\[-8pt]
\multirow{ 2}{*}{$\hat{\mu}$}  & RB & $ 4.1 $ & $2.3 $ & $4.5$ & $2.0$ & $3.8$ & $1.7$ & $2.8$ & $2.1$ & -2.7 & -1.7 \\
& MSE & 8.0 & 3.0 & 8.3 & 3.1 & 8.2 & 2.9 & 8.2 & 3.1 & 8.4 & 3.2 \\
\\[-8pt]
\multirow{2}{*}{$\tilde{\mu}_{0.2}$} & RB & 10.3 & 6.1 & 10.3 & 6.2 & 10.5 & 5.9 & 10.3 & 5.8 & -0.2 & -0.1\\
& MSE & 20.5 & 9.3 & 19.7 & 9.4 & 19.8 & 9.0 & 18.8 & 9.2 & 10.2 & 5.1\\
\\[-8pt]
\multirow{2}{2em}{$\tilde{\mu}_{0.4}$} & RB & 4.6 & 2.6 & 4.7 & 2.7 & 4.4 & 2.5 & 4.3 & 2.6 & -5.5 & -5.6\\
& MSE & 10.3 & 3.8 & 9.8 & 4.1 & 9.7 & 3.7 & 9.2 & 3.9 & 12.8 & 7.3 \\
\\[-8pt]
\multirow{2}{2em}{$\tilde{\mu}_{0.6}$} & RB & 2.5 & 1.6 & 2.8 & 1.6 & 2.7 & 1.4 & 2.6 & 1.7 & -8.1 & -7.1\\
  & MSE & 7.7 & 2.8 & 7.2 & 2.9 & 7.2 & 2.7 & 7.0 & 2.9 & 14.6 & 8.6 \\
\\[-8pt]
\multirow{2}{2em}{$\tilde{\mu}_{0.2,0.4}$} & RB & $ 8.4 $ & $5.5$ & $8.0$ & $4.5$ & $8.2$ & $4.7$ & $7.7$ & $4.7$ & -3.7 & -0.3\\
& MSE & 14.3 & 6.2 & 13.7 & 5.9 & 14.2 & 6.2 & 13.4 & 5.9 & 9.6 & 4.6\\
\\[-8pt]
\multirow{2}{2em}{$\tilde{\mu}_{0.4,0.6}$} & RB & 4.8 & 3.7 & 4.6 & 2.9 & 5.0 & 2.9 & 4.5 & 2.9 & -4.2 & -2.6\\
& MSE & 9.0 & 3.8 & 9.0 & 3.9 & 9.0 & 3.9 & 8.7 & 3.78 & 10.1 & 4.0 \\
\\[-8pt]
\multirow{2}{2em}{${\check{\mu}}$} & RB & 4.2 & 3.7 & 3.9 & 2.7 & 4.3 & 1.9 & 4.8 & 3.7 & -12.4 & -10.1\\
& MSE & 7.8 & 3.7 & 8.5 & 3.5 & 8.0 & 3.3 & 7.6 & 4.1 & 20.9 & 11.6 \\
\\[-8pt]
\multirow{2}{2em}{$\Bar{y}$} & RB & 0.4 & 0.3 & 0.5 & 0 & -0.2 & -0.1 & 0.5 & -0.2 & 0.5 & -0.2 \\
& MSE & 6.3 & 2.5 & 6.3 & 2.6 & 6.1 & 2.6 & 6.7 & 2.4 & 6.7 & 2.4 \\
\\[-8pt]
\multirow{2}{2em}{$\Bar{y}_{obs}$} & RB & 40.5 & 40.8 & 40.5 & 40.1 & 39.2 & 39.2 & 39.3 & 38.8 & -2.1 & -1.3\\
& MSE & 107.7 & 103.4 & 108.3 & 99.9 & 102.1 & 95.7 & 102.5 & 94.0 & 9.7 & 4.8 \\
\\[-8pt]
\multirow{2}{2em}{$\breve{\mu}_{\gamma_0}$} & RB & 0.2 & 0.1 & 0.4 & 0.2 & 0.2 & 0.3 & 0.4 & -0.2 & -1.9 & -1.6\\
& MSE & 6.4 & 2.5 & 6.5 & 2.7 & 6.5 & 2.5 & 6.1 & 2.5 & 9.0 & 3.6 \\
\\[-8pt]
\multirow{2}{2em}{$\breve{\mu}_{\gamma_0/n}$} & RB & 16.6 & 18.4 & 16.7 & 18.5 & 16.4 & 18.2 & 16.1 & 17.2 & -1.9 & -1.6 \\
& MSE & 24.2 & 23.1 & 24.6 & 23.6 & 23.7 & 22.7 & 22.7 & 20.7 & 9.0 & 3.6\\
\\[-8pt]
\multirow{2}{2em}{$\breve{\mu}_{0}$} & RB & 17.5 & 18.8 & 17.5 & 18.9 & 17.2 & 18.6 & 16.9 & 17.6 & -1.9 & -1.6 \\
& MSE & 25.8 & 23.9 & 26.2 & 24.4 & 25.3 & 23.5 & 24.3 & 21.5 & 9.0 & 3.6\\
\\[-8pt]
\end{tabular}}
\label{tab:tbl1}
\end{table}

\begin{table}[t!]
\def~{\hphantom{0}}
\centering
\caption{Standard Error (SE), Coverage Probability (CP) for n=200 and 500; all  CPs are multiplied by 100}{%
\begin{tabular}{lcrrrrrrrrrr}
\\
& &  \multicolumn{10}{c}{Model}\\
&  & \multicolumn{2}{c}{$M_1$} & \multicolumn{2}{c}{$M_2$} & \multicolumn{2}{c}{$M_3$} & \multicolumn{2}{c}{$M_4$} & \multicolumn{2}{c}{$M_5$}   \\
Estimator & Quantity \hspace{0.1in} $n=$ \hspace{-0.1in} & 200 & 500 & 200 & 500 & 200 & 500 & 200 & 500 & 200 & 500  \\
\\[-8pt]
\multirow{ 2}{*}{$\hat{\mu}$}  &  SE & 1.7 & 1.1 & 1.7 & 1.0 & 1.7 & 1.1 & 1.7 & 1.1 & 1.9 & 1.2\\
& CP & 85.9 & 87.3 & 84.2 & 85.8 & 87.2 & 87.4 & 83.9 & 86.2 & 88.0 & 94.0\\
\\[-8pt]
\multirow{2}{*}{$\tilde{\mu}_{0.2}$} &  SE & 2.5 & 1.8 & 2.5 & 1.8 & 2.5 & 1.8 & 2.5 & 1.8 & 2.4 & 1.8\\
& CP & 82.1 & 89.6 & 83.4 & 87.7 & 84.6 & 90.2 & 84.6 & 89.0 & 96.9 & 97.0\\
\\[-8pt]
\multirow{2}{2em}{$\tilde{\mu}_{0.4}$} &  SE & 1.9 & 1.2 & 1.9 & 1.2 & 1.9 & 1.2 & 1.9 & 1.2 & 2.3 & 1.6\\
& CP & 87.2 & 90.3 & 88.5 & 88.3 & 89.7 & 90.6 & 89.3 & 89.0 & 80.6 & 73.3\\
\\[-8pt]
\multirow{2}{2em}{$\tilde{\mu}_{0.6}$} &  SE & 1.7 & 1.1 & 1.7 & 1.1 & 1.7 & 1.1 & 1.7 & 1.1 & 2.2 & 1.5\\
  & CP & 86.7 & 89.2 & 89.1 & 88.5 & 89.7 & 89.9 & 94.7 & 94.9 & 88.0 & 80.0\\
\\[-8pt]
\multirow{2}{2em}{$\tilde{\mu}_{0.2,0.4}$} &  SE & 2.0 & 1.8 & 2.0 & 1.4 & 2.0 & 1.4 & 2.0 & 1.3 & 2.0 & 1.3\\
& CP & 76.7 & 99.1 & 80.5 & 82.8 & 77.9 & 81.3 & 79.3 & 82.4 & 90.0 & 84.0\\
\\[-8pt]
\multirow{2}{2em}{$\tilde{\mu}_{0.4,0.6}$} &  SE & 1.7 & 1.6 & 1.7 & 1.2 & 1.7 & 1.1 & 1.7 & 1.1 & 1.9 & 1.2\\
& CP & 83.6 & 99.8 & 85.0 & 84.7 & 83.4 & 84.0 & 84.7 & 84.6 & 80.0 & 86.0  \\
\\[-8pt]
\multirow{2}{2em}{${\check{{\mu}}}$} &  SE & 1.7 & 1.1 & 1.7 & 1.1 & 1.7 & 1.1 & 1.7 & 1.1 & 2.6 & 2.0\\
& CP & 89.0 & 85.0 & 85.0 & 92.0 & 86.0 & 85.0 & 89.0 & 81.0 & 81.0 & 81.0  \\
\\[-8pt]
\multirow{2}{2em}{$\Bar{y}$} &  SE & 1.6 & 1.0 & 1.6 & 1.0 & 1.6 & 1.0 & 1.6 & 1.0 & 1.6 & 1.0\\
& CP & 88.0 & 89.5 & 90.4 & 88.1 & 89.0 & 88.6 & 86.6 & 90.1 & 86.6 & 90.1\\
\\[-8pt]
\multirow{2}{2em}{$\Bar{y}_{obs}$} &  SE & 2.3 & 1.5 & 2.3 & 1.4  & 2.3 & 1.4 & 2.3 & 1.4 & 2.2 & 1.4\\
& CP & 0.4 & 0.0 & 0.2 & 0.0 & 0.9 & 0.0 & 0.8 & 0.0 & 99.0 & 88.0 \\
\\[-8pt]
\multirow{2}{2em}{$\breve{\mu}_{\gamma_0}$} &  SE & 1.6 & 1.0 & 1.6 & 1.0 & 1.6 & 1.0 & 1.6 & 1.0 & 2.0 & 1.3\\
& CP & 87.4 & 88.6 & 87.8 & 88.4 & 86.7 & 89.3 & 90.4 & 90.8 & 88.2 & 91.4\\
\\[-8pt]
\multirow{2}{2em}{$\breve{\mu}_{\gamma_0/n}$} &  SE & 2.0 & 1.3 & 2.0 & 1.3 & 2.0 & 1.3 & 2.0 & 1.3 & 2.0 & 1.3\\
& CP & 48.0 & 7.2 & 48.1 & 8.7 & 49.7 & 8.6 & 49.5 & 10.9 & 88.2 & 91.4\\
\\[-8pt]
\multirow{2}{2em}{$\breve{\mu}_{0}$} &  SE &  2.0 & 1.3 & 2.0 & 1.3 & 2.0 & 1.3 & 2.0 & 1.3 & 2.0 & 1.3\\
& CP & 44.1 & 6.7 & 43.6 & 7.9 & 46.8 & 7.6 & 46.6 & 9.7 & 88.2 & 91.4\\
\\[-8pt]
\end{tabular}}
\label{tab:tbl2}
\end{table}

\subsection{Using ordinal instrumental variable}
We now assume that the instrument $z$ is ordinal with 3 ordered levels $c_0\prec c_1\prec c_2$ but the distances between these categories  are not known. Let $z^*$ the underlying variable that generates $z$ such that $z=c_k \Leftrightarrow z^*=k$, $k\in \{0, 1, 2\}$, with $p(z=c_k)=p_k$. Assuming $z^*\sim Binomial(2,0.4)$, we consider the hierarchical structure for the full complete data:
\[z\sim \{(c_k, p_k)\}_{k=0}^2,\; u \mid \{z=c_k\} \sim N(k,1),\;  y \mid \{u, z=c_k\} \sim N(E\{y | u,z=c_k\}, 1),\] where \(E\{y | u,z\}=[1+0.5(u-1)^2]I\{z=c_0\}+u^2I\{z=c_1\} + [2+(u-2)^2] I\{z=c_2\}\).

The unconditional mean of $y$ is $\mu=4.13$. Three MNAR models $M_1^*, M_2^*, M_3^*$ and one MAR model  $M^*_4$ were considered, keeping the unconditional rate of missingness at 30\%:\\
$M_1^*$ :\; $\pi=1/\{1+exp(\alpha+\beta u+\gamma y)\}$ with ($\alpha, \beta, \gamma)=(0.4,-0.3,-0.2)$;\\
$M_2^*$ :\; $\pi=1/\{1+exp(\alpha+\beta sin(u)+\gamma y)\}$ with ($\alpha, \beta, \gamma)=(0.2,-0.9,-0.2)$;\\
$M_3^*$ :\; $\pi=1/\{1+exp(\alpha+\beta u^2+\gamma y)\}$ with ($\alpha, \beta, \gamma)=(0.5,-0.2,-0.1)$;\\
$M_4^*$ :\; $\pi=1/\{1+exp(\alpha+\beta u)\}$ with ($\alpha, \beta)=(-0.4,-0.2)$.

Here the finite sample performance of the following estimators of $\mu$ are compared: $\hat{\mu}$ our proposed estimator along with $\tilde{\mu}$ the estimator proposed in Shao \& Wang (2016) with 3 strata; $\Bar{y}$, $\Bar{y}_{obs}$, $\breve{\mu}_{\gamma_0}$, $\breve{\mu}_{\gamma_0/n}$ and $\breve{\mu}_{0}$ as previously described. Because instrument $z$ is not quantitative, the moment-based estimator $\check{\mu}$ is not applied. Results in Table \ref{tab:tbl3} show that $\hat{\mu}$, the scanning-based estimator, performs almost similarly with $\tilde{\mu}$ in terms of bias and MSE. As expected, $\hat{\mu}$ consistently produces higher coverage probabilities of $\mu$ for all  propensity models  relative to $\tilde{\mu}$ (see Table \ref{tab:tbl4}). The other estimators continue to behave as expected. For example, the complete case analysis continues to underperform in terms of bias and MSE, especially when the missing data model deviates substantially from missing completely at random data mechanism.

\begin{table}[t!]
\def~{\hphantom{0}}
\centering
\caption{Relative biases (RB), Mean square errors (MSE) for n=200 and 500; all RBs are multiplied by 100}{%
\begin{tabular}{lcrrrrrrrr}
\\
&  & \multicolumn{8}{c}{Model}\\
&  & \multicolumn{2}{c}{$M_1^*$} & \multicolumn{2}{c}{$M_2^*$} & \multicolumn{2}{c}{$M_3^*$} & \multicolumn{2}{c}{$M_4^*$}     \\
Estimator & Quantity \hspace{0.1in} $n=$ \hspace{-0.1in} & 200 & 500 & 200 & 500 & 200 & 500 & 200 & 500 \\
\\[-8pt]
\multirow{ 2}{*}{$\hat{\mu}$}  & RB &  0.4  & 0.3  & -0.3 & -0.1 & 0.6 & 0.5 & -0.3 & -0.2 \\
& MSE & 0.2 & 0.1 & 0.2 & 0.1 & 0.2 & 0.1 & 0.2 & 0.1 \\
\\[-8pt]
\multirow{2}{*}{$\tilde{\mu}$} & RB & 0.8 & 0.2  & -2.7 & -2.0 & 0.8 & 0.6 & -0.5 & -0.4 \\
& MSE & 0.2 & 0.1 & 0.2 & 0.1 & 0.2 & 0.1 & 0.2 & 0.1 \\
\\[-8pt]
\multirow{2}{2em}{$\Bar{y}$} & RB & 0.0 & -0.1 & -0.1 & 0.0 & -0.2 & -0.2 & -0.2 & -0.2 \\
& MSE & 0.1 & 0.1 & 0.1 & 0.1 & 0.1 & $<\!\!0.1$ & 0.1 & $<\!\!0.1$  \\
\\[-8pt]
\multirow{2}{2em}{$\Bar{y}_{obs}$} & RB & 16.3 & 16.3 & 8.3 & 8.4 & 17.8 & 17.7 & 4.3 & 4.1\\
& MSE & 0.7 & 0.5 & 0.3 & 0.2 & 0.7 & 0.6 & 0.2 & 0.1\\
\\[-8pt]
\multirow{2}{2em}{$\breve{\mu}_{\gamma_0}$} & RB & -0.3 & -0.2 & -1.6 & -0.6 & -0.1 & 0.1 & -0.6 & -0.3\\
& MSE & 0.1 & 0.1 & 0.1 & 0.1 & 0.1 & 0.1 & 0.2 & 0.1\\
\\[-8pt]
\multirow{2}{2em}{$\breve{\mu}_{\gamma_0/n}$} & RB & 4.4 & 4.5 & 6.5 & 7.4 & 1.9 & 2.2  & -0.6 & -0.3\\
& MSE & 0.2 & 0.1 & 0.2 & 0.2 & 0.1 & 0.1 & 0.2 & 0.1\\
\\[-8pt]
\multirow{2}{2em}{$\breve{\mu}_{0}$} & RB & 4.4 & 4.5 & 6.6 & 7.4 & 1.9 & 2.2 & -0.6 & -0.3\\
& MSE & 0.2 & 0.1 & 0.2 & 0.2 & 0.1 & 0.1 & 0.2 & 0.1\\
\\[-8pt]
\end{tabular}}
\label{tab:tbl3}
\end{table}

\begin{table}[t!]
\def~{\hphantom{0}}
\centering
\caption{Standard Error (SE), Coverage Probability (CP) for n=200 and 500; all  CPs are multiplied by 100}{%
\begin{tabular}{lcrrrrrrrr}
\\
&  & \multicolumn{8}{c}{Model}\\
&  & \multicolumn{2}{c}{$M_1^*$} & \multicolumn{2}{c}{$M_2^*$} & \multicolumn{2}{c}{$M_3^*$} & \multicolumn{2}{c}{$M_4^*$}     \\
Estimator & Quantity \hspace{0.1in} $n=$ \hspace{-0.1in} & 200 & 500 & 200 & 500 & 200 & 500 & 200 & 500 \\
\\[-8pt]
\multirow{ 2}{*}{$\hat{\mu}$}  & SE &  0.3  & 0.2  & 0.3 & 0.2 & 0.3 & 0.2& 0.3 & 0.2 \\
& CP & 95.0 & 94.3 & 94.4 & 95.2 & 94.5 & 93.1 & 93.5 & 94.5\\
\\[-8pt]
\multirow{2}{*}{$\tilde{\mu}$} & SE & 0.3  & 0.2  & 0.3 & 0.2 & 0.3 & 0.2 & 0.3 & 0.2  \\
& CP & 90.8 & 90.2 & 86.7 & 87.4 & 90.1 & 91.1 & 91.0 & 94.0\\
\\[-8pt]
\multirow{2}{2em}{$\Bar{y}$} & SE & 0.2 & 0.2 & 0.2 & 0.2 & 0.2 & 0.2 & 0.2 & 0.2 \\
& CP & 92.5 & 92.8 & 92.9 & 93.0 & 92.6 & 92.5& 92.6 & 92.5  \\
\\[-8pt]
\multirow{2}{2em}{$\Bar{y}_{obs}$} & SE & 0.3 & 0.2 & 0.3 & 0.2 & 0.3 & 0.2 & 0.3 & 0.2 \\
& CP & 44.0 & 8.4 & 83.0 & 59.7 & 36.1 & 3.4 & 91.6 & 88.1\\
\\[-8pt]
\multirow{2}{2em}{$\breve{\mu}_{\gamma_0}$} & SE & 0.3 & 0.2 & 0.3 & 0.2 & 0.3 & 0.2 & 0.3 & 0.2 \\
& CP & 91.4 & 93.4 & 93.4 & 92.3 & 92.9 & 93.5 & 91.9 & 92.5\\
\\[-8pt]
\multirow{2}{2em}{$\breve{\mu}_{\gamma_0/n}$} & SE & 0.3 & 0.2 & 0.3 & 0.2 & 0.3 & 0.2 & 0.3 & 0.2 \\
& CP & 90.7 & 81.5 & 87.4 & 68.8 & 93.2 & 92.7 & 91.9 & 92.5\\
\\[-8pt]
\multirow{2}{2em}{$\breve{\mu}_{0}$} & SE & 0.3 & 0.2 & 0.3 & 0.2 & 0.3 & 0.2 & 0.3 & 0.2 \\
& CP & 90.6 & 81.5 & 87.3 & 68.6 & 93.2 & 92.7 & 91.9 & 92.5 \\
\\[-8pt]
\end{tabular}}
\label{tab:tbl4}
\end{table}

\section{Income panel data}

We reanalyze income data that were generated from the Korean Labor and Income Panel Study.
Detailed account  of this survey and previous related analyses focusing on missingness can be found elsewhere (e.g., Wang et al, 2014;\nocite{wang:2014} and Shao \& Wang, 2016). In this survey, 2506 regular wage earners were interviewed for their 2006 monthly income regarded as the response $y$, in addition to their prior year income,   gender (male versus female),  age (years) and education level (high school and below versus above high school). For our analysis, the 2005 monthly income is treated as the $u$ variable. There were about 35\% missing data on the $y$ variable, but the other variables were fully observed. Our analysis focuses on the overall average income in 2006, accounting for potentially nonrandom missingness. Indeed, missingness from self-reported income is well known to be informative in that both lower and higher ends of the income spectrum oftentimes do not report their income (Schenker et al., 2006)\nocite{schenker:2006}.

Several estimators of the overall 2006 monthly income using age as instrument were generated alongside the complete case analysis, the naive analysis relying on single thresholds. Additionally, sensitivity analysis varying the instruments was also conducted. Results are reported in Table \ref{tab:real-data-anlayis}. The naive estimates are generally less precise than the proposed scanning-based estimates, demonstrating the need to accumulate information across the instrument. Surprisingly, the estimate under the complete case analysis has a reduced standard error albeit being  substantially larger than the other estimates.

\begin{table}[h]
\def~{\hphantom{0}}
\centering
\caption{Estimates and standard errors of average monthly income in 2006, treating Age as a continuous instrumental variable, and additional sensitivity analysis results for various instruments}{%
\begin{tabular}{lccc}
 \\
Method & Additional Instruments & Estimates (SE)  \\
\\
Complete case analysis & no instrument & 205.7 (2.8)\\
\\[-8pt]
 \multirow{4}{*}{Scanning across age} & -- & 189.7 (2.3) \\
&  gender & 185.5 (3.2)  \\
&  education & 185.9 (3.7)  \\
& gender, education & 184.7 (0.6)  \\
\\[-8pt]
 \multirow{4}{*}{Single age threshold (37)} & -- & 190.1 (5.1) \\
&  gender & 184.9 (3.1)  \\
&  education & 183.9 (3.6)  \\
&  gender, education & 184.7 (4.0)  \\
\\[-8pt]
 \multirow{4}{*}{Multiple age thresholds (32,44)} & -- & 193.3 (11.5) \\
& gender & 187.4 (4.8)  \\
& education & 186.2 (4.6)  \\
& gender, education & 186.9 (5.6)  \\
\\[-8pt]
 \multirow{4}{*}{1st moment of age} & -- & 190.3 (9.3) \\
& gender & 183.7 (8.4)  \\
& education & 187.3 (4.9)  \\
& gender, education & 185.5 (4.3)  \\
\\[-8pt]
\end{tabular}}
\label{tab:real-data-anlayis}
\end{table}

\section{Discussion}
\label{sec:Discussion}

The proposed scanning approach is broadly applicable to instruments with an intrinsic order, including quantitative and ordinal categorical variables. Because it naturally embeds the underlying order into the estimation process, it typically will outperform methods that ignore such information in efficiency. An interesting feature of the methodology is that it can be combined with multiple instruments not requiring scanning (e.g, categorical instruments). It can also be applied to multiple instruments jointly requiring scanning, but the computational cost for such an undertaking is often onerous. An alternative approach that shares this  feature to some extent is the moment-based estimator, which is only applicable to quantitative instruments.

Computationally, an advantage of dichotomization is that it eliminates the need for a second step generalized moment estimator of $\gamma$. Obviously for a more complicated propensity  model such as
\(\pi(y,u)=\left[ 1+ \exp\{g(u)+\gamma_1 y+\gamma_2 y^2\} \right]^{-1},\) the use of dichotomization may not be helpful with identification. In which case,  it may be necessarily to entertain continuous thresholding with multiple cutpoints to ensure identifiability of $\gamma_1$ and $\gamma_2$.

Because $z$ serves as a missing data instrument for $y$, it must meet the basic requirement of being related to $y$, and being excluded from the propensity model. Consequently, the law of $f(y\mid x)$ must be related to $z$. As such, it may be of interest to evaluate the effect of $y$ on $z$. We may adopt, as a byproduct of the scanning approach, the quantity \[r=\left\{{\int_{\Xi^*} p_{\xi}(1-p_{\xi})d\xi}\right\}^{-1}{\int_{\Xi^*} p_{\xi}(1-p_{\xi})\{\mbox{E}\{y \mid z\leq \xi\}- \mbox{E}\{y \mid z > \xi\}\}   d\xi},\]
as a nonparametric measure of the effect of instrument $z$ on $y$. Obviously, if $y$ and $z$ are non associated $r=0$. Here $\mbox{E}\{y \mid z\in A\}$ is the mean of $y$ restricted to the population stratified by $A$ of $z$.  To estimate $\mbox{E}\{y | z\in A \}$ for a stratum $A$ of $z$, one may use \(\{n_{_A}\}^{-1}\sum_{i=1}^n {I\{z_i \in A\}\delta_i y_i/\{\pi(y_i,u_i;\hat{g}_{\hat{\gamma}_\xi},\hat{\gamma}_\xi)\}}, \xi \in {\Xi}^{*}\), with $n_{_A}=\sum_{i=1}^n I\{z_i \in A\}$. Thus for $\xi^* \in {\Xi}^{*}$, estimators of $\mbox{E}\{y |z\leq \xi^* \}$ and $\mbox{E}\{y |z> \xi^* \}$ then follow.

The proposed method relies essentially on a readily available instrument, which is often unrealistic in practice. Some technical non-design based guidelines for the choice of viable instruments can be found elsewhere (e.g., Shao \& Wang, 2016; and  Zhao \emph{et al.}, 2021\nocite{zhao:2021:sufficient}). Additionally,  the proposed method uses the Kernel-based estimation of the unknown function $g(\cdot)$, which may prove computationally difficult in the case of high-dimensional auxiliary variables, $u$. In such settings, any sufficient dimension reduction techniques may be entertained (e.g., Cook and Li, 2002;\nocite{cook:li:2002}  Wang \emph{et al.}, 2020; and Zhao \emph{et al.}, 2021).

\section*{Acknowledgements}

The authors are grateful to Professor Jun Shao for sharing the income panel data.  This research is supported by NSF grant DMS-1916339 and NIH/NIDCR grant R03DE027108.

\appendix
\section*{Appendix}

\begin{center}{\it Key regularity conditions}\end{center}

\begin{condition}\label{A1}
The space $\Gamma$ of $\gamma$ and $\Theta$ that of $\theta$ are all compact.
\end{condition}

\begin{condition}
The kernel $K(u)$ has bounded derivatives of order d and satisfies $\int K(u)du=1$, and has zero moments of orders up to $m-1$ and nonzero m-th order.
\end{condition}
\begin{condition}
The true function of $g(u)$ is continuously differentiable and bounded on an open
set containing the support of u.
\end{condition}
\begin{condition}
The moment $E\{exp(4\gamma y)\}$ is finite and the function $E\{exp(4\gamma y)|u\}f (u)$ is
bounded, where $f(u)$ is the marginal density of u.
\end{condition}
\begin{condition}
The threshold dependent bandwidth $h_{\xi,n}$ is such that  as $n \to \infty$, $\sup_{\xi \in \Xi^*} h_{\xi,n}\to 0$, $ \inf_{\xi \in \Xi^*}  nh_{\xi,n}^{p_u} \to \infty$, $\inf_{\xi \in \Xi^*}  n^{1/2}h_{\xi,n}^{p_u+2d}/\log n \to \infty$ and $\sup_{\xi \in \Xi^*} n h_{\xi,n}^{2m} \to 0$, where $p_u$ is the dimension of u.
\end{condition}

\begin{condition}
For any threshold $\xi \in \Xi$, there is a vector of the functional $G^\xi(y, u, \delta, \omega)$ which is linear in $\omega = (\omega_1, \omega_2)^T$ and such that:
    \begin{enumerate}
        \item for small enough $\parallel \omega-\omega_0 \parallel$, $\parallel \tilde{m}^\xi(y, u, \delta, \omega, \gamma_0)-\tilde{m}^\xi(y, u, \delta, \omega_0, \gamma_0)-G^\xi(y, u, \delta, \omega-\omega_0)\parallel \leq b^\xi(y,u,\delta)(\parallel \omega-\omega_0 \parallel)^2$, where $\tilde{m}^\xi(y, u, \delta, \omega^\xi, \gamma)=I(z \leq \xi)[\delta\{1+exp(\gamma y)\omega_1(u)/\omega_2(u)\}-1]$,  $\omega_0= [E(1-\delta \mid u), E\{\delta exp(\gamma_0 y)\mid u\}]^T$ and $\sup_{\xi \in \Xi}E(b^\xi(y,u,\delta)) < \infty$;
        \item $\parallel G^\xi(y, u, \delta, \omega) \parallel \leq c^\xi(y,u,\delta)\parallel w \parallel$ and $\sup_{\xi \in \Xi}E(c^\xi(y,u,\delta)^2) \leq \infty$;
        \item There exists an almost everywhere continuous function $\nu^\xi(u)$ with $\sup_{\xi \in \Xi}\int \parallel \nu^\xi(u) \parallel du \leq \infty$, $E(G^\xi(y, u, \delta, \omega))=\int \nu^\xi(u) \omega(u) du$, for all $\parallel \omega \parallel < \infty$ and $\sup_{\xi \in \Xi}E\{\sup_{\parallel \zeta \parallel \leq \epsilon}\parallel \nu^\xi(u+\zeta)\parallel^4\}<\infty$, for some $\epsilon>0$.
 \end{enumerate}

\end{condition}

\begin{condition}
 For small enough $\parallel \omega-\omega_0 \parallel$, $\tilde{m}^\xi(y, u, \delta, \omega, \gamma)$ is continuously differentiable
in $\gamma$ for any threshold $\xi \in \Xi$, in a neighbourhood of $\gamma_0$, and there is $k^\xi(y, u, \delta)$ with $sup_{\xi \in \Xi} E\{k^\xi(y, u, \delta)\} < \infty$ such that $\parallel \bigtriangledown_\gamma \tilde{m}^\xi(y, u, \delta, \omega, \gamma)-\bigtriangledown_\gamma \tilde{m}^\xi(y, u, \delta, \omega_0, \gamma_0)\parallel \leq k^\xi(y, u, \delta)(\parallel \gamma - \gamma_0 \parallel^\epsilon + \parallel \omega-\omega_0 \parallel^\epsilon)$, for any $\epsilon>0$, $\Gamma^\xi=E\{\bigtriangledown_\gamma \tilde{m}^\xi(y, u, \delta, \omega_0, \gamma_0)\}$ exists and of full rank.
\end{condition}

\begin{condition}
 Assume that ${V}(\tilde{\theta}_\xi)$ the variance of $\,\tilde{\theta}_\xi$ is uniformly positive definite over $\xi \in {\Xi^*}$;  that is  $\inf_{\xi \in {\Xi^*}} \mbox{eigmin}{V}(\tilde{\theta}_\xi)>0$,  where $\mbox{eigmin}(.)$ denotes the smallest eigenvalue.
 \end{condition}

A sufficient condition for the previous assumption to hold is that the limiting point of ${V}(\tilde{\theta}_\xi)$ is uniformly positive definite over $(\xi, \theta) \in {\Xi^*}\times \Theta$;  that is  $\inf_{(\xi, \theta) \in {\Xi^*}\times \Theta} \mbox{eigmin}({V}_\xi({\theta}))>0$, where $\mbox{eigmin}(.)$ denotes the smallest eigenvalue.


\bibliographystyle{unsrtnat}
\bibliography{ms}  

\begin{thebibliography}{28}
\providecommand{\natexlab}[1]{#1}
\providecommand{\url}[1]{\texttt{#1}}
\expandafter\ifx\csname urlstyle\endcsname\relax
  \providecommand{\doi}[1]{doi: #1}\else
  \providecommand{\doi}{doi: \begingroup \urlstyle{rm}\Url}\fi

\bibitem[Molenberghs et~al.(1997)Molenberghs, Kenward, and
  Lesaffre]{MOLENBERGHS:KENWARD:LESAFFRE:1997}
G.~Molenberghs, M.~G. Kenward, and E.~Lesaffre.
\newblock {The analysis of longitudinal ordinal data with nonrandom drop-out}.
\newblock \emph{Biometrika}, 84\penalty0 (1):\penalty0 33--44, 03 1997.
\newblock ISSN 0006-3444.
\newblock \doi{10.1093/biomet/84.1.33}.
\newblock URL \url{https://doi.org/10.1093/biomet/84.1.33}.

\bibitem[Robins et~al.(1994)Robins, Rotnitzky, and
  Zhao]{robins:rotnitzky:zhao:1994}
James~M. Robins, Andrea Rotnitzky, and Lue~Ping Zhao.
\newblock Estimation of regression coefficients when some regressors are not
  always observed.
\newblock \emph{Journal of the American Statistical Association}, 89\penalty0
  (427):\penalty0 846--866, 1994.
\newblock \doi{10.1080/01621459.1994.10476818}.
\newblock URL \url{https://doi.org/10.1080/01621459.1994.10476818}.

\bibitem[Tsiatis and Davidian(2007)]{tsiatis2007}
Anastasios~A. Tsiatis and Marie Davidian.
\newblock Comment: Demystifying double robustness: A comparison of alternative
  strategies for estimating a population mean from incomplete data.
\newblock \emph{Statist. Sci.}, 22\penalty0 (4):\penalty0 569--573, 11 2007.

\bibitem[Rubin(1987)]{rubin:1987}
{Donald B.} Rubin.
\newblock \emph{Multiple Imputation for nonresponse in surveys}.
\newblock Wiley series in probability and mathematical statistics. Wiley, New
  York, NY, 1987.
\newblock ISBN 978-0-471-08705-2.

\bibitem[Little and Rubin(2002)]{Litt:Rubi:stat:2002}
Roderick J.~A. Little and Donald~B. Rubin.
\newblock \emph{Statistical Analysis with Missing Data}.
\newblock John Wiley \& Sons, 2002.
\newblock ISBN 0-471-18386-5.

\bibitem[Kim and Shao(2021)]{kim:shao:2021}
Jae~Kwang Kim and Jun Shao.
\newblock \emph{Statistical methods for handling incomplete data}.
\newblock CRC press, 2021.

\bibitem[Linero(2017)]{linero:2017}
Antonio~R Linero.
\newblock Bayesian nonparametric analysis of longitudinal studies in the
  presence of informative missingness.
\newblock \emph{Biometrika}, 104\penalty0 (2):\penalty0 327--341, 2017.

\bibitem[Yang et~al.(2019)Yang, Wang, and Ding]{yang:2019}
S~Yang, L~Wang, and P~Ding.
\newblock {Causal inference with confounders missing not at random}.
\newblock \emph{Biometrika}, 106\penalty0 (4):\penalty0 875--888, 09 2019.
\newblock ISSN 0006-3444.
\newblock \doi{10.1093/biomet/asz048}.
\newblock URL \url{https://doi.org/10.1093/biomet/asz048}.

\bibitem[Scharfstein et~al.(1999)Scharfstein, Rotnitzky, and
  Robins]{Scha:Rotn:Robi:adju:1999}
Daniel~O. Scharfstein, Andrea Rotnitzky, and James~M. Robins.
\newblock Adjusting for nonignorable drop-out using semiparametric nonresponse
  models ({C}/{R}: P1121-1146).
\newblock \emph{Journal of the American Statistical Association}, 94:\penalty0
  1096--1120, 1999.

\bibitem[Rotnitzky et~al.(2001)Rotnitzky, Scharfstein, Su, and
  Robins]{Rotn:Scha:Su:Robi:meth:2001}
Andrea Rotnitzky, Daniel Scharfstein, Ting-Li Su, and James Robins.
\newblock Methods for conducting sensitivity analysis of trials with
  potentially nonignorable competing causes of censoring.
\newblock \emph{Biometrics}, 57\penalty0 (1):\penalty0 103--113, 2001.

\bibitem[Zhao and Shao(2015)]{zhao:shao:2015}
Jiwei Zhao and Jun Shao.
\newblock Semiparametric pseudo-likelihoods in generalized linear models with
  nonignorable missing data.
\newblock \emph{Journal of the American Statistical Association}, 110\penalty0
  (512):\penalty0 1577--1590, 2015.

\bibitem[Todem et~al.(2010)Todem, Fine, and Peng]{Todem:2009}
David Todem, Jason~P. Fine, and Limin Peng.
\newblock A global sensitivity test for evaluating statistical hypotheses with
  nonidentifiable models.
\newblock \emph{Biometrics}, 66:\penalty0 558--566, 2010.

\bibitem[Wu and Carroll(1988)]{wu:carroll:1988}
Margaret~C. Wu and Raymond~J. Carroll.
\newblock Estimation and comparison of changes in the presence of informative
  right censoring by modeling the censoring process.
\newblock \emph{Biometrics}, 44\penalty0 (1):\penalty0 175--188, 1988.
\newblock ISSN 0006341X, 15410420.

\bibitem[Troxel et~al.(1998b)Troxel, Lipsitz, and
  Harrington]{Trox:Lips:Harr:marg:1998}
Andrea~B. Troxel, Stuart~R. Lipsitz, and David~P. Harrington.
\newblock Marginal models for the analysis of longitudinal measurements with
  nonignorable non-monotone missing data.
\newblock \emph{Biometrika}, 85:\penalty0 661--672, 1998b.

\bibitem[Wang et~al.(2014)Wang, Shao, and Kim]{wang:2014}
Sheng Wang, Jun Shao, and Jae~Kwang Kim.
\newblock An instrumental variable approach for identification and estimation
  with nonignorable nonresponse.
\newblock \emph{Statistica Sinica}, pages 1097--1116, 2014.

\bibitem[Shao and Wang(2016)]{shao:wang:2016}
Jun Shao and Lei Wang.
\newblock {Semiparametric inverse propensity weighting for nonignorable missing
  data}.
\newblock \emph{Biometrika}, 103\penalty0 (1):\penalty0 175--187, 01 2016.
\newblock ISSN 0006-3444.
\newblock \doi{10.1093/biomet/asv071}.
\newblock URL \url{https://doi.org/10.1093/biomet/asv071}.

\bibitem[Tchetgen~Tchetgen and Wirth(2017)]{tchetgen:wirth:2017}
Eric~J. Tchetgen~Tchetgen and Kathleen~E. Wirth.
\newblock A general instrumental variable framework for regression analysis
  with outcome missing not at random.
\newblock \emph{Biometrics}, 73\penalty0 (4):\penalty0 1123--1131, 2017.

\bibitem[Kim and Yu(2011)]{kim:yu:2011}
Jae~Kwang Kim and Cindy~Long Yu.
\newblock A semiparametric estimation of mean functionals with nonignorable
  missing data.
\newblock \emph{Journal of the American Statistical Association}, 106\penalty0
  (493):\penalty0 157--165, 2011.

\bibitem[Altman(1998)]{Altman:1998}
Douglas~G. Altman.
\newblock Categorizing continuous variables.
\newblock In \emph{{Encyclopedia of Biostatistics, P. Armitage and T. Colton
  (eds)}}, pages 563--567. Chichester, U.K.: Wiley, 1998.
\newblock ISBN 9780470011812.

\bibitem[Peng and Fine(2008)]{peng:fine:2008}
L~Peng and JP~Fine.
\newblock Nonparametric tests for continuous covariate effects with multistate
  survival data.
\newblock \emph{Biometrics}, 64\penalty0 (4):\penalty0 1080--1089, 2008.

\bibitem[Royston et~al.(2006)Royston, Altman, and
  Sauerbrei]{Royston:Altman:Sauerbrei:2006}
Patrick Royston, Douglas~G. Altman, and Willi Sauerbrei.
\newblock Dichotomizing continuous predictors in multiple regression: a bad
  idea.
\newblock \emph{Statistics in Medicine}, 25\penalty0 (1):\penalty0 127--141,
  2006.
\newblock \doi{https://doi.org/10.1002/sim.2331}.
\newblock URL \url{https://onlinelibrary.wiley.com/doi/abs/10.1002/sim.2331}.

\bibitem[Farewell et~al.(2004)Farewell, Tom, and Royston]{farewell:2004}
Vern~T Farewell, Brian D.~M Tom, and Patrick Royston.
\newblock The impact of dichotomization on the efficiency of testing for an
  interaction effect in exponential family models.
\newblock \emph{Journal of the American Statistical Association}, 99\penalty0
  (467):\penalty0 822--831, 2004.
\newblock \doi{10.1198/016214504000001169}.
\newblock URL \url{https://doi.org/10.1198/016214504000001169}.

\bibitem[Khan and Powell(2001)]{khan:2001}
Shakeeb Khan and James~L Powell.
\newblock Two-step estimation of semiparametric censored regression models.
\newblock \emph{Journal of Econometrics}, 103\penalty0 (1):\penalty0 73--110,
  2001.
\newblock ISSN 0304-4076.
\newblock \doi{https://doi.org/10.1016/S0304-4076(01)00040-9}.
\newblock URL
  \url{https://www.sciencedirect.com/science/article/pii/S0304407601000409}.
\newblock Studies in estimation and testing.

\bibitem[Dudley(1978)]{dudley:1978}
R.~M. Dudley.
\newblock {Central Limit Theorems for Empirical Measures}.
\newblock \emph{The Annals of Probability}, 6\penalty0 (6):\penalty0 899 --
  929, 1978.
\newblock \doi{10.1214/aop/1176995384}.
\newblock URL \url{https://doi.org/10.1214/aop/1176995384}.

\bibitem[Gasser and M{\"u}ller(1979)]{gasser:muller:1979}
Theo Gasser and Hans-Georg M{\"u}ller.
\newblock Kernel estimation of regression functions.
\newblock In Th. Gasser and M.~Rosenblatt, editors, \emph{Smoothing Techniques
  for Curve Estimation}, pages 23--68, Berlin, Heidelberg, 1979. Springer
  Berlin Heidelberg.
\newblock ISBN 978-3-540-38475-5.

\bibitem[Schenker et~al.(2006)Schenker, Raghunathan, Chiu, Makuc, Zhang, and
  Cohen]{schenker:2006}
Nathaniel Schenker, Trivellore~E Raghunathan, Pei-Lu Chiu, Diane~M Makuc,
  Guangyu Zhang, and Alan~J Cohen.
\newblock Multiple imputation of missing income data in the national health
  interview survey.
\newblock \emph{Journal of the American Statistical Association}, 101\penalty0
  (475):\penalty0 924--933, 2006.

\bibitem[Zhao et~al.(2021)Zhao, Wang, and Shao]{zhao:2021:sufficient}
Puying Zhao, Lei Wang, and Jun Shao.
\newblock Sufficient dimension reduction and instrument search for data with
  nonignorable nonresponse.
\newblock \emph{Bernoulli}, 27\penalty0 (2):\penalty0 930--945, 2021.

\bibitem[Cook and Li(2002)]{cook:li:2002}
R.Dennis Cook and Bing Li.
\newblock {Dimension reduction for conditional mean in regression}.
\newblock \emph{The Annals of Statistics}, 30\penalty0 (2):\penalty0 455 --
  474, 2002.
\newblock \doi{10.1214/aos/1021379861}.
\newblock URL \url{https://doi.org/10.1214/aos/1021379861}.

\end{thebibliography}


\begin{thebibliography}{7}
\bibitem[1]{SW}  Shao, J.,  Wang, L. (2016). Semiparametric inverse propensity weighting for nonignorable missing data. Biometrika, 103(1), 175-187.

\bibitem[2]{Newey} Newey, W. K.,  McFadden, D. (1994). Large sample estimation and hypothesis testing. Handbook of econometrics, 4, 2111-2245.

\bibitem[3]{Dudley} Dudley, R. M. (1978). Central limit theorems for empirical measures. The Annals of Probability, 899-929.

\end{thebibliography}






\newpage
\begin{center}
\textbf{\large Supplemental Material: Nonparametric scanning for nonrandom missing data with continuous instrumental variables}
\end{center}
\section*{Proofs Of The Technical Results}

\textbf{ Proof of Theorem \ref{thm1}} \\
By Theorem 1 of Shao and Wang (2016) , we have the pointwise consistency of $\tilde{\gamma}_\xi$ to $\gamma_{0\xi}$. So we only need to show the uniform convergence of the process $\tilde{\gamma}_\xi$ in $\xi$. For that we first recall the following definition and relevant theorem of stochastic equicontinuity from Newey \& McFadden \(1994\). A sequence of random function $Q_n(\theta)$ is stochastic equicontinuious if 
\[\forall \epsilon>0,  \lim\limits_{\delta \to 0} \limsup\limits_{n\to\infty} P[\sup\limits_{|\theta-\theta'|<\delta} |Q_n(\theta)-Q_n(\theta')|>\epsilon]=0\]
And then by Lemma 2.8, Newey \& McFadden (1994), for compact $\Theta$ and continuous $Q_0(\theta)$, $\sup\limits_{\theta \in \Theta} |\tilde{Q}_n(\theta)-Q_0(\theta)| \overset{P}{\to} 0$ if and only if $\tilde{Q}_n(\theta) \overset{P}{\to} Q_0(\theta)$ $\forall \theta \in \Theta$ and $\tilde{Q}_n(\theta)$ is stochastic equicontinuous.  \\

Next, we consider $\tilde{Q}_n(\xi)=|\tilde{\gamma}_\xi-\gamma_{0\xi}|$. Here the threshold variable $\xi \in \Xi$ and to prove the uniform convergence we only consider a dense subset and the rational part $\Xi^R$  of this compact set $\Xi$. And we define, $\Xi_\delta^R= \{\xi,\xi'\in \Xi^R$ such that $|\xi-\xi'|<\delta\}$. Now,

\begin{math}
\lim\limits_{\delta \to 0} \limsup\limits_{n\to\infty} P[\sup\limits_{|\xi-\xi'|<\delta} ||\tilde{\gamma}_\xi - \gamma_{0\xi}|-|\tilde{\gamma}_{\xi'} - \gamma_{0\xi'}||>\epsilon]\\
\leq \lim\limits_{\delta \to 0} \limsup\limits_{n\to\infty} P[\sup\limits_{\Xi_\delta^R} (|\tilde{\gamma}_\xi - \gamma_{0\xi}|+|\tilde{\gamma}_{\xi'} - \gamma_{0\xi'}|)>\epsilon]\\
\leq  \lim\limits_{\delta \to 0} \limsup\limits_{n\to\infty} P[\exists \text{ at least one} (\xi,\xi')\in {\Xi_\delta^R}\times {\Xi_\delta^R}, \text{ such that } (|\tilde{\gamma}_\xi - \gamma_{0\xi}|+|\tilde{\gamma}_{\xi'} - \gamma_{0\xi'}|)>\epsilon ]\\
= \lim\limits_{\delta \to 0} \limsup\limits_{n\to\infty}P[\bigcup\limits_{\xi,\xi' \in \Xi_\delta^R} \{|\tilde{\gamma}_\xi - \gamma_{0\xi}|+|\tilde{\gamma}_{\xi'} - \gamma_{0\xi'}|>\epsilon\}]\\
\leq  \lim\limits_{\delta \to 0} \limsup\limits_{n\to\infty} \sum\limits_{\xi,\xi' \in \Xi_\delta^R} P[|\tilde{\gamma}_\xi - \gamma_{0\xi}|+|\tilde{\gamma}_{\xi'} - \gamma_{0\xi'}|>\epsilon]\\
\leq \lim\limits_{\delta \to 0} \limsup\limits_{n\to\infty} \sum\limits_{\xi,\xi' \in \Xi_\delta^R} \{ P(|\tilde{\gamma}_\xi - \gamma_{0\xi}|>\epsilon/2)+P(|\tilde{\gamma}_{\xi'} - \gamma_{0\xi'}|>\epsilon/2)\}\\
\end{math}
Here by applying dominated convergence theorem, we can take the $\limsup$ inside the sum and then the pointwise convergence of $\tilde{\gamma}_\xi$ to $\gamma_{0\xi}$ gives the required stochastic continuity of $\tilde{Q}_n(\xi)$.\\
Hence, we establish the uniform consistency of $\tilde{Q}_n(\xi)$, i.e. $\sup\limits_{\xi \in \Xi} |\tilde{\gamma}_\xi - \gamma_{0\xi}| \overset{P}{\to} 0$.

\textbf{ Proof of Theorem \ref{thm3}} \\
Here, for the simplicity of the proof we consider $\theta=\mu=E(Y) $. Following  Dudley (1978), we will prove this in two steps: i) we will show that any finite dimensional distribution of the stochastic process $n^{1/2}(\tilde{\mu}_\xi - \mu)$  converges to a multivariate normal distribution; and ii) we will show its stochastic equicontinuity to prove the tightness of the asymptotic Gaussian process.
Now, we prove the first part using the Cramér–Wold theorem, i.e. for any finite $d \in \{1,2,...\}$, $t_1 X_{n_{\xi_1}}+t_2 X_{n_{\xi_2}}+...+t_d X_{n_{\xi_d}}$ is convergent in distribution to $t_1 X_{\xi_1}+ t_2 X_{\xi_2}+...+t_d X_{\xi_d}$, where $X_{n_{\xi_k}} = n^{1/2}(\tilde{\mu}_{\xi_k} - \mu)$ and ($X_{\xi_1}, X_{\xi_2},...,X_{\xi_d}$) follow some multivariate Gaussian distribution.\\
We consider the following expansion as in Shao and Wang (2016) of identically and independently distributed random variables with zero mean and finite variance .\\
For $t_1,t_2, \dots,t_d\in \Re^d$, \\
\begin{math}
t_1 n^{1/2}(\tilde{\mu}_{\xi_1} - \mu)+t_2 n^{1/2}(\tilde{\mu}_{\xi_2} - \mu)+\dots+t_d n^{1/2}(\tilde{\mu}_{\xi_d} - \mu)\\
=n^{1/2} \sum_{i=1}^n \sum_{j=1}^d t_j \left\{\frac{\delta_i (y_i - v_i)}{\pi(y_i,u_i,g_{\gamma_{0\xi_j}},\gamma_{0\xi_j})} + (v_i - \mu)+ \zeta_i (\xi_j)\right\}+o_p(1) \overset{CLT}{\to} \sum_{j=1}^d t_j R_{\xi_j}\\
\end{math}\\
where $v_i = E(y_i/u_i)$ and $R_{\xi_j}  \sim N\left(0, var\left\{\frac{\delta (y - v)}{\pi(y,u,g_{\gamma_{0\xi_j}},\gamma_{0\xi_j})} + (v-\mu)+\zeta(\xi_j)\right\}\right)$ for $j\in \{1,2,\dots,d\}$.\\

Hence by Cramér–Wold theorem, any finite dimensional distribution of $n^{1/2}(\tilde{\mu}_\xi - \mu)$ is converging to a multivariate normal distribution. \\
Now, to establish the stochastic equicontinuity, we need to show\\
\begin{math}
    \lim\limits_{\delta \downarrow 0} \limsup\limits_{n\to\infty} P\left[\sup\limits_{|\xi-\xi'|<\delta} |n^{1/2}(\tilde{\mu}_\xi - \mu) - n^{1/2}(\tilde{\mu}_{\xi'} - \mu)|>\epsilon\right]=0
\end{math}\\
Lets denote, $T_n(\delta)=\sup\limits_{|\xi-\xi'|<\delta} |n^{1/2}(\tilde{\mu}_\xi - \mu) - n^{1/2}(\tilde{\mu}_{\xi'} - \mu)|$.\\
If we can show that the limits in the stochastic equicontinuity conditions are interchangeable, then for any fixed $n$ and for all $\delta<\delta_0$, where $\delta_0= \min\{|t-t'|, t,t' \in [z_{(i-1)},z_{(i)}], i=2,3,\dots,n\}$, $\tilde{\mu}_\xi=\tilde{\mu}_{\xi'}$ with probability 1 where $|\xi-\xi'|<\delta$ and hence $T_n(\delta)=0$. Hence the stochastic equicontinuity holds and the limiting Gaussian process is tight.\\
So we only need to show that the two limits are indeed interchangeable. By the Moore-Osgood Theorem on exchanging limits, the sufficient condition for that is: a) $\limsup_{n \to \infty} T_n(\delta)=T(\delta)$ uniformly on $\delta \in \Delta, \Delta \subset \mathcal{R} $ and b) $\lim_{\delta \downarrow 0}T_n(\delta)$ converges pointwise on $n\in \mathcal{N}$. While (b) holds by the above mentioned argument with the specific choice of $\delta_0$ conditioned on the instrument z; (a) can be proved by applying the continuous mapping theorem with the previously mentioned convergence of any finite dimensional distribution of $n^{1/2}(\tilde{\mu}_\xi - \mu)$ to multivariate Gaussian distribution.\\

\textbf{ Proof of Theorem \ref{thm4}} \\
Under the regularity condition where bootstraping gives a valid consistent variance estimator, we consider $W(\xi)=\frac{1}{\hat{V}(\hat{\mu}_\xi)} \overset{a.s.}{\to}\frac{1}{V(\hat{\mu}_\xi)}= w(\xi)$ for any fixed $\xi$. Now, \\
\begin{math}
n^{1/2}(\hat{\mu}-\mu) \overset{a}{=} n^{1/2} \frac{\int_\Xi (\hat{\mu}_\xi -\mu) w(\xi)d\xi}{\int_\Xi w(\xi)d\xi}
\overset{a}{=} n^{1/2} \sum_{i=1}^n  \underbrace{\frac{\int_\Xi \left(\frac{\delta_i (y_i - v_i)}{\pi(y_i,u_i,g_{\gamma_{0\xi}},\gamma_{0\xi})} + (v_i - \mu)+ \zeta_i (\xi)\right)w(\xi)d\xi}{\int_\Xi w(\xi)d\xi}}_{\text{iid terms with mean 0}}\\
\end{math}
Here we assume that $E\left( \left[\int_\Xi \left\{\frac{\delta_i (y_i - v_i)}{\pi(y_i,u_i,g_{\gamma_{0\xi}},\gamma_{0\xi})} + (v_i - \mu_i)+ \zeta_i (\xi)\right\}w(\xi)d\xi\right]^2 \right) <\infty$\\
Then by applying CLT we get the asymptotic normality of our proposed estimator.

\section*{Additional Simulations - Using moments of the instrumental variable in estimating equation}

For continuous instruments, it was suggested in Shao and Wang (2016) to use moment-based estimating equations as an alternative approach to discretizing the instrument; for example replacing the indicator function in the estimating equations by any $\ell$th moment of $z$ and using up to $L$  estimating equations,  $\ell=1,2,...,L$. The exact specification of this approach is given by Equation (8) in the main paper.
A real difficulty in practice is to determine the optimal number of moments for the estimating equation without knowing the true data generating mechanism. This is critical as the performance of the moment-based approach may depend on the number of moments and the underlying structure of the data. In general,
setting $L$ equal to the dimension of the tilting parameter $\gamma$ works reasonably well in practice. But for some other missing data settings, this method may not work well regardless of the moments of the instrument being used.
Using a simulation study, we show that for the zero inflated instrumental variable, the moment-based approach does not work well in contrast to our proposed scanning method.

Here we consider the following incomplete data structure:\\
{\bf Full data generating mechanism}
\begin{math}\\
z=\Delta W_1+(1-\Delta) W_2, \text{ where } \Delta \sim {Bernoulli(0.7)}, W_1=\delta_{\{0\}}, W_2\sim N(0,0.1)\\
u|\{z\} \sim N\left(z-5,1\right) \text{ and } y|\{u,z\} \sim N\left( u^2+0.1z, 1 \right)
\end{math}\\
Averaging with an MC approximation gives $\mu=26$ for $ E(y)$ the  unconditional mean of $y$.\\
{\bf Missing data generating mechanism}
We have considered the following propensity models, keeping the unconditional rate of missingness at 30$\%$:
\begin{enumerate}
\item $M_1^*$: $\pi_i=1/\{1+exp(\alpha+\beta u_i+\gamma y_i)\}$ with ($\alpha, \beta, \gamma)=(9.8,-0.5,-0.6)$;
\item $M_2^*$: $\pi_i=1/\{1+exp(\alpha+\beta exp(u_i)+\gamma y_i)\}$ with ($\alpha, \beta, \gamma)=(11.8,-0.5,-0.6)$;
\end{enumerate}
In Tables \ref{tab:tbl5} and \ref{tab:tbl6}, along with those of $\hat{\mu}$ our proposed estimator, $\Bar{y}$, $\Bar{y}_{obs}$, $\breve{\mu}_{\gamma_0}$, $\breve{\mu}_{\gamma_0/n}$ and $\breve{\mu}_{0}$ as described in the main paper, the finite sample performance of the following three estimators of $\mu$ are compared:
\begin{enumerate}
\item $\check{\mu}_{1}$, the estimator using the 1st moment of $z$ in the estimating equation, $L=1$;
\item $\check{\mu}_{1,2}$, the estimator using the 1st two moments of $z$ in the estimating equation, $L=2$;
\item $\check{\mu}_{1,2,3}$, the estimator using the 1st three moments of $z$ in the estimating equation, $L=3$.
\end{enumerate}

Results show that the moment-based approach clearly fails here, contrarily to the scanning approach which works comparatively better for both propensity models.

\FloatBarrier
\begin{table}[t!]
\centering
\def~{\hphantom{0}}
\caption{Relative biases (RB), Mean square errors (MSE) for n=200 and 500; all RBs are multiplied by 100}{%
\begin{tabular}{lcrrrr}
&  & \multicolumn{4}{c}{Model}\\
&  & \multicolumn{2}{c}{$M_1^*$} & \multicolumn{2}{c}{$M_2^*$} \\
Estimator & Quantity \hspace{0.1in} $n=$ \hspace{-0.1in} & 200 & 500 & 200 & 500\\
\\[-8pt]
\multirow{ 2}{*}{$\hat{\mu}$}  & RB &  -2.9  & -2.5  & -3.3 & -3.0 \\
 & MSE & 2.2 & 1.3 & 2.4 & 1.5 \\
 \\[-8pt]
\multirow{2}{*}{$\check{\mu}_1$} & RB & -11.5  & -10.7  & -11.3 & -11.3 \\
& MSE & 12.5 & 10.6 & 12.2 & 11.0 \\
\\[-8pt]
\multirow{2}{*}{$\check{\mu}_{1,2}$} & RB & -8.9 & -6.7  & 8.8 & -6.9 \\
& MSE & 10.1 & 6.6 & 9.8 & 6.8 \\
\\[-8pt]
\multirow{2}{*}{$\check{\mu}_{1,2,3}$} & RB & -9.2  & -6.8  & -9.5 & -7.2 \\
& MSE & 10.7  & 6.9 & 10.9 & 7.0 \\
\\[-8pt]
\multirow{2}{2em}{$\Bar{y}$} & RB & -0.1 & 0.0 & 0.0 & 0.0 \\
& MSE & 0.8 & 0.3 & 0.8 & 0.3 \\
\\[-8pt]
\multirow{2}{2em}{$\Bar{y}_{obs}$} & RB & 16.9 & 17.0 & 16.9 & 16.9 \\
& MSE & 20.3 & 19.9 & 20.2 & 19.8 \\
\\[-8pt]
\multirow{2}{2em}{$\breve{\mu}_{\gamma_0}$} & RB & -4.2 & -3.4 & -3.9 & -3.3 \\
& MSE & 2.9 & 1.7 & 2.7 & 1.7 \\
\\[-8pt]
\multirow{2}{2em}{$\breve{\mu}_{\gamma_0/n}$} & RB & -6.1 & -4.8 & -6.2 & -5.2 \\
& MSE & 4.2 & 2.5 & 4.2 & 2.7 \\
\\[-8pt]
\multirow{2}{2em}{$\breve{\mu}_{0}$} & RB & -6.1 & -4.8 & -6.2 & -5.2 \\
& MSE & 4.2 & 2.5 & 4.3 & 2.7 \\
\\[-8pt]
\end{tabular}}
\label{tab:tbl5}
\end{table}
\FloatBarrier

\FloatBarrier
\begin{table}[b!]
\centering
\def~{\hphantom{0}}
\caption{ Standard Error (SE), Coverage Probability (CP) for n=200 and 500; all  CPs are multiplied by 100}{%
\begin{tabular}{lccccc}
&  & \multicolumn{4}{c}{Model}\\
&  & \multicolumn{2}{c}{$M_1^*$} & \multicolumn{2}{c}{$M_2^*$} \\
Estimator & Quantity \hspace{0.1in} $n=$ \hspace{-0.1in} & 200 & 500 & 200 & 500\\
 \\[-8pt]
\multirow{ 2}{*}{$\hat{\mu}$}  & SE &  0.7  & 0.4  & 0.7 & 0.5 \\
 & CP & 64.0 & 52.3 & 63.4 & 45.1 \\
 \\[-8pt]
\multirow{2}{*}{$\breve{\mu}_1$} & SE & 1.4  & 1.3  & 1.3 & 1.3  \\
& CP & 44.9 & 43.3 & 37.6 & 33.1 \\
\\[-8pt]
\multirow{2}{*}{$\breve{\mu}_{1,2}$} & SE & 1.4  & 1.3  & 1.3 & 1.2 \\
& CP & 57.5 & 72.8 & 51.4 & 64.6 \\
\\[-8pt]
\multirow{2}{*}{$\breve{\mu}_{1,2,3}$} & SE & 1.3  & 1.2  & 1.3 & 1.2 \\
& CP & 49.6 & 66.7 & 41.6 & 61.1 \\
\\[-8pt]
\multirow{2}{2em}{$\Bar{y}$} & SE & 0.5 & 0.3 & 0.5 & 0.3 \\
& CP &85.1 & 84.3 & 84.3 & 84.8 \\
\\[-8pt]
\multirow{2}{2em}{$\Bar{y}_{obs}$} & SE & 0.7 & 0.5 & 0.7 & 0.4 \\
& CP & 0.9 & 0.0 & 0.0 & 0.0 \\
\\[-8pt]
\multirow{2}{2em}{$\breve{\mu}_{\gamma_0}$} & SE & 0.7 & 0.5 & 0.7 & 0.5 \\
& CP & 52.3 & 37.6 & 54.9 & 41.8 \\
\\[-8pt]
\multirow{2}{2em}{$\breve{\mu}_{\gamma_0/n}$} & SE & 0.6 & 0.4 & 0.7 & 0.4 \\
& CP & 33.4 & 19.6 & 31.2 & 16.6 \\
\\[-8pt]
\multirow{2}{2em}{$\breve{\mu}_{0}$} & SE & 0.6 & 0.4 & 0.6 & 0.4 \\
& CP & 32.9 & 19.6 & 30.5 & 16.6 \\
\\[-8pt]
\end{tabular}}
\label{tab:tbl6}
\end{table}
\FloatBarrier

\renewcommand\refname{References For The Supplementary Material}

\end{document}